\documentclass[preprint,12pt,review]{elsarticle}

\usepackage{amssymb}
\usepackage{amsmath}
\usepackage[all]{xy}

\journal{Physics Letters B}

\begin{document}

\begin{frontmatter}



\title{Distributions for tau neutrino interactions observed through the decay $\tau\rightarrow \mu\nu_\tau\bar\nu_\mu$}

\author{I. Alikhanov\corref{cor1}}

\cortext[cor1]{{\it Email address: {\tt ialspbu@gmail.com}}}

\address{Institute for Nuclear Research of the Russian Academy of Sciences, Moscow 117312, Russia\\
Research Institute for Applied Mathematics and Automation, Nalchik 360000, Russia}

\author{E.~A.~Paschos\corref{cor2}}

\cortext[cor2]{{\it Email address: {\tt paschos@physik.uni-dortmund.de}}}
\address{Department of Physics, TU Dortmund, D-44221 Dortmund, Germany}

\begin{abstract}
We investigate the problem of identifying $\nu_\tau$-induced reactions
in large neutrino telescopes. We concentrate on events with tracks and
showers where the respective energies $E_\mu$ and $E_X$ are measured separately. Then we compute analytically and numerically event
distributions in two variables $r_l = E_X/E_l$   and
$y'=E_{\text{X}}/(E_{\mu}+E_{X})$. We find that the $y'$-distribution
is especially useful because in the $\nu_\tau$-induced reactions the
distribution has a minimum at $y'=0.75$ and then increases as
$y'\rightarrow1$. This is different in $\nu_\mu$-induced reactions
where the $y'$-distribution decreases monotonically. The results are
demonstrated with figures where one can estimate the required
sensitivity of the experiments. Another attractive property is that in
several ratios the neutrino flux factorizes and drops out. We hope the
results of this article will be useful for searches of tau
(anti)neutrinos of energies above 100 GeV to several TeV in the present
and planned large volume neutrino telescopes.
\end{abstract}


\end{frontmatter}

\section{Introduction}
The discovery of ultra-high energy cosmic neutrinos~\cite{Aartsen:2013jdh,Aartsen:2014gkd} opens a unique window to test neutrino properties at very high energies~\cite{Beacom:2002vi,Beacom:2003nh}.
In these experiments the lepton flavor composition of the neutrino flux is essential for understanding mechanisms of generation, propagation and interactions of the neutrinos. There are many running experiments such as ANITA~\cite{Hoover:2010qt}, ARA~\cite{Chen:2009ra}, ANTARES~\cite{antares} and IceCube~~\cite{Aartsen:2013jdh,Aartsen:2014gkd}, as well as the next generation deep water neutrino telescopes KM3NeT~\cite{km3net} and NT1000~\cite{baikal} (Lake Baikal) which will provide results for cosmic neutrinos. Of primary interest is the flavor content of the cosmic flux, because even though the initial mechanisms produce mostly muon and electron neutrinos, oscillations over cosmological distances generate a substantial $\nu_\tau$ component.

The detection of $\nu_\tau$'s in the telescopes has attracted special interest~\cite{Cowen:2007} and the various event topologies have been classified. An analysis of the IceCube data using the double pulse algorithm did not establish any $\nu_\tau$ component at energies 214~TeV to 72~PeV~\cite{Aartsen:2015dlt}. A second possibility to identify tau neutrinos uses the decay mode $\tau\rightarrow\mu\nu_{\tau}\bar\nu_{\mu}$ and separates the events into showers and tracks~\cite{Cowen:2007,Aartsen:2015ivb,Aartsen:2015knd}. In this approach the observable quantities are $E_{\mu}$ and $E_{\text{shower}}$. The IceCube Collaboration classified their data into showers and tracks and uses a likelihood function in order to obtain the best fit of the events as a function of the sum $E_{\mu}+E_{\text{shower}}$~\cite{Aartsen:2015ivb,Aartsen:2015knd}. They combined the results with the astronomical flux and deduced its flavor content with a large uncertainty. A more detailed analysis is possible when the energies of the muons, $E_{\mu}$ , and of the hadronic showers, $E_{\text{shower}}$, are measured separately in the same event. The distributions in the two variables are sensitive to the $\nu_\tau$ component.

The purpose of this article is to provide explicit distributions in ratios of two variables and point out quantities which are less dependent on the shape of the incident fluxes and on the quark distribution functions. Reactions \eqref{eq:recmu} and \eqref{eq:rectau} differ because the values for the energy $E_{\mu}$ are lower for reaction \eqref{eq:rectau} and the two reactions have different angular distributions. Our results are computed analytically and it is possible to adopt them to the experimental situations.

\section{Analytic results\label{sec2}}
We wish to distinguish the two CC reactions:

\begin{equation}
     \nu_{\mu}+N\rightarrow \mu+X, \label{eq:recmu}
\end{equation}

\begin{equation}
 \xymatrix{
       \hskip 3.2cm&\nu_{\tau}+N\rightarrow \tau+X   \ar [rd]&  \\
                                &&\hskip1.cm\mu+\nu_{\tau}+\bar\nu_{\mu},}\label{eq:rectau}
\end{equation}
where $X$ denotes the hadronic final state. The hadrons will initiate showers in the detector and the muons will produce single tracks, as illustrated schematically in Fig.~\ref{fig1}. In both cases the energy distributions in the hadronic showers are the same, but the $E_{\mu}$ distributions differ because of the intervening $\tau$-decay. The leptonic $\tau$-decay produces two invisible neutrinos. It remains to quantify how much lower is the muon energy in the second reaction. For this purpose we study distributions in the two dimensionless variables 

\begin{equation}
 r_l=\frac{E_X}{E_{l}}\,\,\,\,\,(0\leq r_l\leq \infty),\label{eq:ratio_def1}
\end{equation}
with $l$ denoting $\mu$ or $\tau$ leptons and the visible inelasticity,

\begin{equation}
 y'=\frac{E_X}{E_\mu+E_X}=\frac{r_\mu}{1+r_\mu} \,\,\,\,\,(0\leq y'\leq 1).\label{eq:ratio_def}
\end{equation}

In this investigation the neutrino energies are very high; they range above 100~GeV to many 10.0 TeV, where the momentum transfer squared, $Q^2$, is small relative to $M_W^2$ and we use the four-point interaction. In addition at this energy range the CC cross section behaves as $\sigma\sim const\cdot E_{\nu}$ which follows from Bjorken scaling. In the following we use the parton model cross sections where both $\nu_\mu N$ and $\nu_\tau N$ processes are calculated with the same quark distribution functions. We shall show later that many results are independent of the quark distribution functions and also independent of the neutrino/antineutrino fluxes.

In the above approach the differential cross section for muon neutrinos is

\begin{equation}
\frac{d\sigma^{(\mu)}}{dr_\mu}=\frac{2G_F^2ME_\nu}{\pi}\frac{1}{(1+r_\mu)^2}\left[Q+\bar Q\frac{1}{(1+r_\mu)^2}\right],
\label{eq:sigmup5}
\end{equation}
where $\overset{\text{{\tiny (}}-\text{{\tiny )}}}{Q}=\int\limits_0^1 x \overset{\text{{\tiny (}}-\text{{\tiny )}}}{q}(x)dx$ are the integrals over quark or antiquark distribution functions.

Reaction~\eqref{eq:rectau} is a two step process. first the $\tau$-lepton is produced and then it decays. The cross section differential on the muon energy is a convolution over the energies of the intermediate $\tau$-lepton
\begin{equation}
\frac{d\sigma^{(\tau)}}{dE_\mu}=\int\limits_{E_{\mu}}^{E_{\nu}}\frac{d\sigma}{dE_\tau}\frac{1}{\Gamma_{\tau\rightarrow\text{all}}}\frac{d\Gamma_{\tau\rightarrow \mu+\nu_{\tau}+\bar\nu_{\mu}}}{dE_{\mu}}dE_{\tau},\label{eq:sigmup6}
\end{equation}
where $\dfrac{d\sigma}{dE_\tau}$ is the cross section for reaction~\eqref{eq:rectau} differential to the $\tau$-lepton energy, $\Gamma_{\tau\rightarrow\text{all}}$ is the total decay width of the tau, $\dfrac{d\Gamma_{\tau\rightarrow \mu+\nu_{\tau}+\bar\nu_{\mu}}}{dE_{\mu}}$ is the differential decay width of the tau with respect to the muon energy in the reference frame where $\tau$ moves with energy~$E_{\tau}$.

This two step process is analogous to the parton model calculations~\cite{Bjorken:1969in}. The initial $\nu_\tau$--nucleon interaction is the source of a beam of $\tau$-leptons with a known "lepton distribution function". It is followed by the known fragmentation of the $\tau$'s into muons. The result of a straightforward calculation is

\begin{eqnarray}
\frac{d\sigma^{(\tau)}}{dr_\mu}=\frac{G^2 ME_{\nu}}{9\pi}\frac{\text{Br}}{r_\mu^5(1+r_\mu)^2} \left\{96
   \left(Q+\bar Q\right)r_\mu+\left(306\,Q+198 \,\bar Q\right)r_\mu^2\right.\nonumber\\\left.+\left(275\, Q+113\,
   \bar Q\right)r_\mu^3+\left(16\,Q+10\,\bar Q\right)r_\mu^4-\left(49\,Q-5 \,\bar Q\right)r_\mu^5\right.\nonumber\\\left.-6 \left(1+r_\mu\right)^2\left[16
   (Q+\bar Q)+9 (3\,Q+\bar Q)\,r_\mu-5\, Q\,r_\mu^3\right]\ln (1+r_\mu) \right\}\label{eq:distrmm},
\end{eqnarray}

where Br is the branching ratio for the decay $\tau\rightarrow \mu+\nu_{\tau}+\bar\nu_\mu$.
Equations~\eqref{eq:sigmup5} and~\eqref{eq:distrmm} are for neutrino-induced reactions; for antineutrinos we interchange the quark and antiquarks distribution functions $Q\leftrightarrow\bar Q$. We also notice that $r_\mu$ and $E_{\nu}$ are independent variables with the neutrino energy appearing as an overall factor. Thus when we multiply the cross sections with the flux factors the integrals over neutrino fluxes factorize.

Let $J_{\nu_l}(E_{\nu})$ be the flux factor of flavor $l$ measured in units $1/(\text{GeV}\,\text{cm}^2\,\text{s}\,\text{sr})$. The fluxes of the extraterrestrial neutrinos and antineutrinos are expected to be same:

\begin{equation}
J_{\nu_l}(E_\nu)=J_{\bar\nu_l}(E_{\nu})=J_{\nu}(E_{\nu})\,\,\,\,\,(l=\mu,\tau).
\end{equation} 
They may be different for various flavors and we present later figures for various compositions. 
Then the yield averaged over the fluxes is 

\begin{eqnarray}
\frac{dN^{(\nu)}}{dr_\mu}=T\,\Omega\, n \int\limits_{E_{\nu\,\text{min}}}^{E_{\nu\,\text{max}}}\left[\frac{d\sigma^{(\mu)}}{dr_\mu}J_{\nu_{\mu}}(E_\nu)+\frac{d\sigma^{(\tau)}}{dr_\mu}J_{\nu_{\tau}}(E_{\nu})\right]dE_{\nu},\label{eq:events3}
\end{eqnarray}
where $T$ is the time of exposure, $\Omega$ is the solid angle of coverage, $n$ is the effective number of nucleons in the target, $E_{\nu\,\text{min}}$ and $E_{\nu\,\text{max}}$ define the range of neutrino energies investigated. 

For illustration we consider the case where the $\nu_\mu$ and $\nu_\tau$ fluxes are equal. We sum over neutrino and antineutrino interactions and normalize to the total number of events. Note that a molecule of ice and/or water neutrino detectors consists of  10  protons  and  8 neutrons being close to isoscalar, and we do not treat the protons and neutrons separately~\cite{Chen:2013dza}. The final result reads

\begin{eqnarray}
\frac{1}{N^{(\nu+\bar\nu)}}\frac{dN^{(\nu+\bar\nu)}}{dr_\mu}=\frac{1}{12(1+\text{Br}) r_\mu^5 (1+r_\mu)^4}\nonumber\\ \times\left[3 \text{Br} \left(5 r_\mu^3-36 r_\mu-32\right) (1+r_\mu)^4 \log
   (1+r_\mu)\right.\nonumber\\\left.-\text{Br} \left(22
   r_\mu^5-13 r_\mu^4-194 r_\mu^3-252 r_\mu^2-96r_\mu\right)(1+r_\mu)^2 +9 r_\mu^5 \left(r_\mu^2+2 r_\mu+2\right)\right].\label{eq:events34}
\end{eqnarray}

It is remarkable that the $r_\mu$-distribution~\eqref{eq:events34} depends neither  on the incoming (anti)neutrino energy nor on the quark distribution functions. The formalism presented above can be easily adopted to cases where the flavor compositions of the neutrino flux are different. Figure~\ref{fig3} shows the $r_\mu$-distributions for the $\nu_{\mu}:\nu_{\tau}=1:0$ and $\nu_{\mu}:\nu_{\tau}=1:1$ compositions. For the curves and the rest of the article we use $\text{Br}=0.17$~\cite{pdg}.

As seen in Fig.~\ref{fig3}, the $r_\mu$-distribution is not dramatically affected by the presence of equal fraction of tau (anti)neutrinos in the flux. There is only $\sim10\%$ change at relatively large values of $r_\mu$. At the same time, in this region the relative number of expected neutrino events drops down by orders of magnitude and it would be difficult to resolve experimentally the difference in the $r_\mu$-distributions. A reason for such a behavior is that the energy spectrum of muons coming from taus decaying in flight has no longer the distinct peak at $E_{\mu}=E_{\tau}/2$ as in the tau rest frame, but  becomes a relatively smooth function of $E_{\mu}$ without peculiarities (see Fig.~\ref{fig2}). The tau neutrinos may more clearly manifest themselves if we concentrate all events in an interval which does not stretch as $r_\mu$ from zero to infinity but is much smaller. This may be achieved, for example, by using the visible inelasticity $y'$ which varies from zero to unity. 

Proceeding in the same way as we described above, one can calculate the $y'$-distribution for the case $\nu_{\mu}:\nu_{\tau}=1:1$ by adding the events originating from $\nu_{\mu}$ and $\nu_{\tau}$ interactions and obtain

\begin{eqnarray}
\frac{1}{N^{(\nu+\bar\nu)}}\frac{dN^{(\nu+\bar\nu)}}{dy'}=\frac{1}{12(1+\text{Br}) y'^5}\nonumber\\ \times\left[\text{Br}\left(3y'^5-3y'^4+14y'^3-132y'^2+96y'\right)\right.\nonumber\\\left.-3 \text{Br} \left(y'^3-24y'^2+60y'-32\right)\log
   \left(1-y'\right)+9 y'^5 \left(y'^2-2 y'+2\right)\right].\label{eq:events34f}
\end{eqnarray}

For other flavor compositions of the neutrino flux we can also calculate the visible inelasticity distribution. We computed the cases $\nu_{\mu}:\nu_{\tau}=1:0$, $\nu_{\mu}:\nu_{\tau}=1:1$ and $\nu_{\mu}:\nu_{\tau}=0:1$ shown in Fig.~\ref{fig4}. One can see that in the case of equal fluxes of muon and tau neutrinos there is a dramatic increase in the number of events as $y'\rightarrow1$. We also present the situation of a purely $\nu_{\tau}+\bar\nu_{\tau}$ flux which is an idealized model of the so-called upward-going neutrinos. This occurs when the neutrinos travel through the earth and the $\nu_e$ and $\nu_{\mu}$-components are absorbed thus only tau neutrinos survive at the detector.
Details of the calculations are given in reference~\cite{our_paper}.

\section{Summary}

The fact that many models predict a substantial component of tau
neutrinos in the extraterrestrial neutrino flux raises the challenge
of obtaining direct or indirect evidence for their presence.

The large volume neutrino telescopes may have the capability to
provide such evidence, but this will require considerable experimental
effort. The challenge is to determine which distributions, accessible
to experiments, are sensitive to this effect.

\begin{figure}
\centering
\includegraphics[width=14cm]{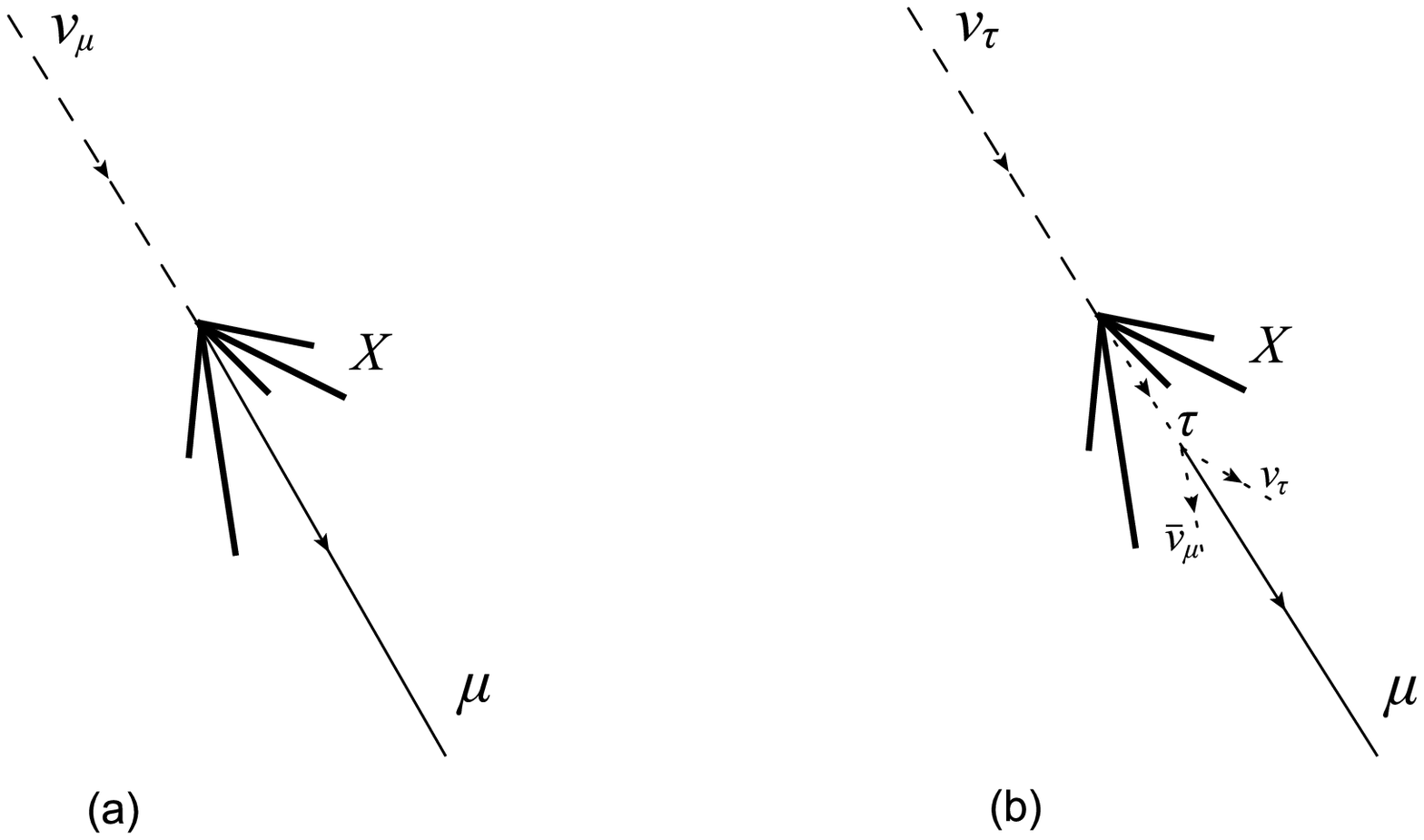}
\caption{}
\label{fig1}
\end{figure}

\begin{figure}
\centering
\includegraphics[width=14cm]{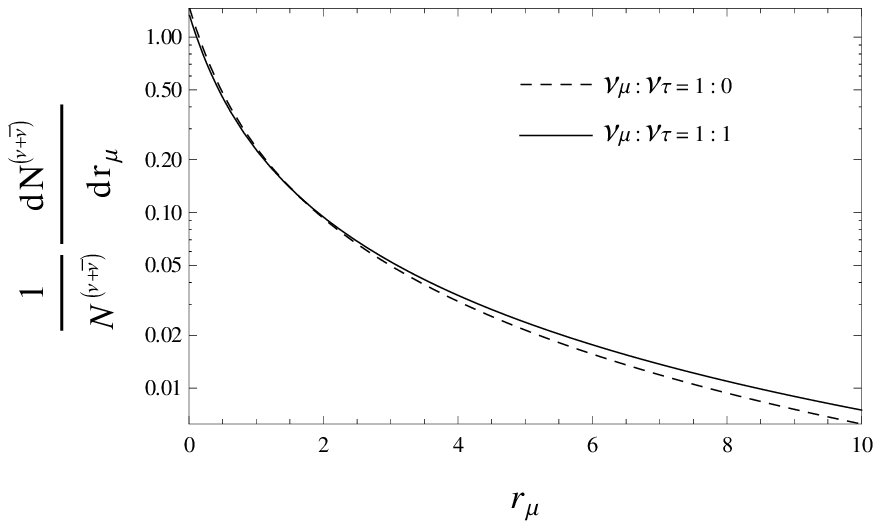}
\caption{}
\label{fig3}
\end{figure}

\begin{figure}
\centering
\includegraphics[width=14cm]{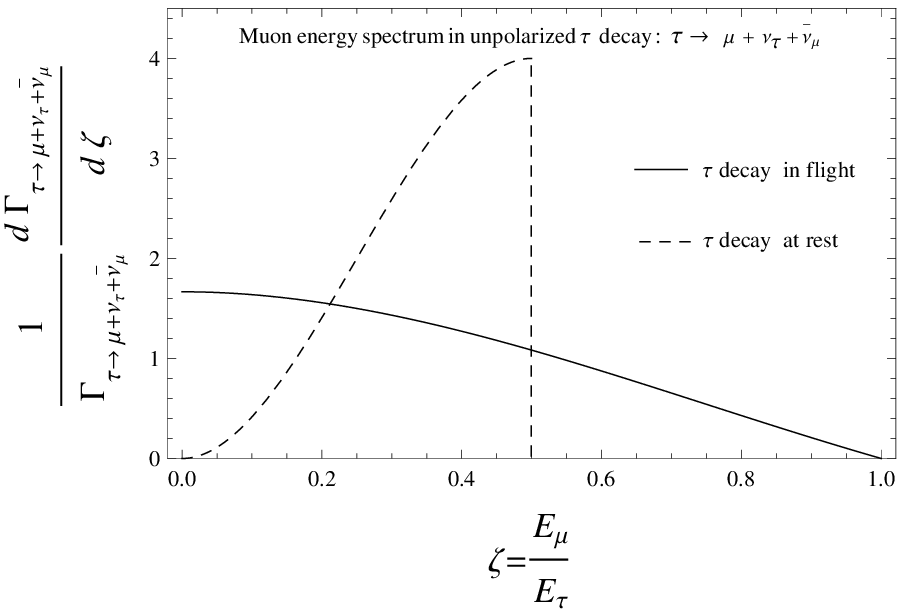}
\caption{}
\label{fig2}
\end{figure}

\begin{figure}
\centering
\includegraphics[width=14cm]{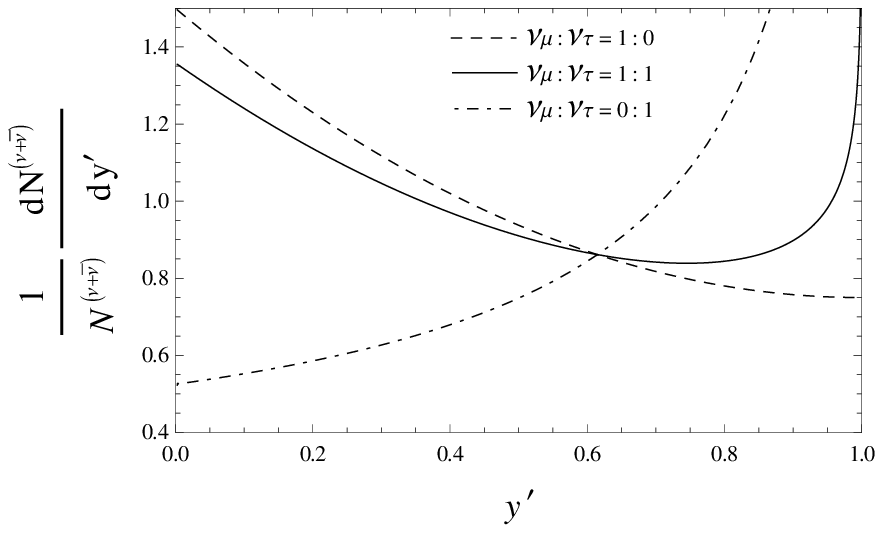}
\caption{}
\label{fig4}
\end{figure}

The analysis of this work concerns only those neutrino events which contain simultaneously a hadronic shower and a muon illustrated in Fig.~\ref{fig1}. We computed the corresponding event distributions for reactions~\eqref{eq:recmu} and~\eqref{eq:rectau} as functions of the dimensionless variables $r_l = E_X/E_l$   and $y'=E_{\text{X}}/(E_{\mu}+E_{X})$. We adopted the quark-parton model and the
four-fermion interaction (no $W$-propagator) working well for $E_\nu\lesssim10$ TeV. However we expect our results not to change significantly at higher neutrino energies, since we consider ratios of the cross sections in which case the $W$-propagator corrections cancel out to an extent. We emphasize an advantage of the approximations adopted in this article that the quark distribution functions factorize and drop out in ratios as well. The same is true for the neutrino flux which makes possible to collect events at lower energies. We find that more $\nu_\tau$-induced events populate the region $y' > 0.60$ as shown in Fig.~\ref{fig4}.
For example, the presence of equal fraction of tau and muon
neutrinos in the flux produces events with muons which have a minimum
at $y'=0.75$ and an excess of events as $y'\rightarrow1$; in contrast the muon
neutrino induced events are decreasing monotonically as $y'\rightarrow1$. For the
$\nu_{\mu}:\nu_{\tau}=1:1$ flavor composition, the ratio of the number of events above $y'=0.60$ to that below is 0.55, while for muon neutrinos only this ratio equals to 0.46. Thus there is a difference between the ratios of the order of $\text{Br}(\tau\rightarrow \mu+\nu_{\tau}+\bar\nu_\mu)\sim20\%$. It will therefore be necessary and efficient to rely on the difference in the shapes of the two curves. In other words, this
$\nu_\tau$-component should lead to an increase of the number of slow or 
stopping muons in the detector in comparison to the case of a purely
$\nu_\mu$-flux. To make a significant distinction between these curves we estimated that~$\gtrsim10^2$ neutrino events with final muons are needed. Assuming the best fit for the diffuse astrophysical neutrino flux~\cite{Aartsen:2014muf}, this number of events will be collected in a few years in the IceCube and other large volume neutrino telescopes. In addition the region $y'=1$ ($E_{\mu}=0$) must be excluded from the analysis because the background from CC interactions of $\nu_e$'s and $\nu_\tau$'s and NC interactions is very large. Nevertheless, one can approach this point as close as it is allowed by the detector muon energy threshold. Likewise, the case $y'=0$ with no visible cascade can be imitated by atmospheric muons penetrating into the detector. The contribution from the atmospheric muons and neutrinos can be rejected by defining an outer veto volume in the detector and by observing extensive air showers on coincidence with neutrino events. A full detector simulation including these backgrounds with Monte Carlo generators can describe in detail the behavior of the proposed distributions for each experiment.

\vskip 0.5cm
{\bf Acknowledgements}
\vskip 0.5cm

One of us (I.A.) acknowledges DAAD support through
the funding program {\it Research Stays for University Academics and
Scientists} and hospitality at TU Dortmund where this work was
initiated. I.A. also thanks H.~P\"as for an invitation to attend weekly meetings of his research group. I.A. was supported in part by the Program for Basic Research of the Presidium of the Russian Academy of Sciences {\it Fundamental Properties
of Matter and Astrophysics}.


\newpage

{\bf Figure captions}
\vskip 0.5 cm 

{\bf Fig. 1:} Schematic illustration of neutrino events with muons in the final state: (a) charged current muon neutrino scattering; (b) charged current tau neutrino scattering with the subsequent decay $\tau\rightarrow\mu\nu_\tau\bar\nu_\mu$. The final hadron state (shower) is denoted by $X$.

\vskip 0.5 cm

{\bf Fig. 2:} Production spectra as a function of the shower-to-muon energy ratio ($r_\mu=E_X/E_{\mu}$). We show two flavor compositions of the neutrino flux: $\nu_{\mu}:\nu_{\tau}=1:0$ (dashed) and $\nu_{\mu}:\nu_{\tau}=1:1$ (solid).

\vskip 0.5 cm

{\bf Fig. 3:} Energy spectrum of muons produced in the leptonic decays of unpolarized taus at rest (dashed) and in flight (solid). Note that at rest $E_{\tau}=m_\tau$.

\vskip 0.5 cm

{\bf Fig. 4:} Production spectra as a function of the visible inelasticity ($y'=E_X/(E_\mu+E_X)$). We show three flavor compositions of the neutrino flux: $\nu_{\mu}:\nu_{\tau}=1:0$ (dashed), $\nu_{\mu}:\nu_{\tau}=1:1$ (solid) and $\nu_{\mu}:\nu_{\tau}=0:1$ (dash-dotted).


\end{document}